\documentclass[
reprint,
superscriptaddress,
amsmath,
amssymb,
aps,
prl
]{revtex4-1}

\usepackage{graphicx}
\usepackage{xcolor} 
\usepackage{multirow}
\usepackage{amssymb}
\usepackage{amsmath}
\usepackage{array}
\usepackage{capt-of}
{

\begin{document}

\title{Intelligent Transportation Systems to Mitigate Road Traffic Congestion}

\author{Nizar Hamadeh}
\author{Ali Karouni}
\author{Zeinab Farhat}
\author{Hussein El Ghor}
\affiliation{LENS Laboratory, Lebanese University, Faculty of Technology, Department of Computer and Communication Engineer, Saida, Lebanon.}

\author{Mohamad El Ghor}
\affiliation{Research Assistant, LENS Laboratory.}

\author{Israa Katea}

\date{\today}
\begin{abstract}

Intelligent transport systems have efficiently and effectively proved themselves in settling up the problem of traffic congestion around the world. The multi-agent based transportation system is one of the most important intelligent transport systems, which represents an interaction among the neighbouring vehicles, drivers, roads, infrastructure and vehicles. In this paper, two traffic management models have been created to mitigate congestion and to ensure that emergency vehicles arrive as quickly as possible. A tool-chain SUMO-JADE is employed to create a microscopic simulation symbolizing the interactions of traffic. The simulation model has showed a significant reduction of at least 50\% in the average time delay and thus a real improvement in the entire journey time.

\end{abstract}

\maketitle

\section{Introduction}
Traffic congestion is one of the most common problems around the world \cite{1}. Recent initiative studies, proved that road traffic congestion is one of the main causes for air and noise pollution. Recently, the number of cars has increased dramatically, as a result of the increasing population, which has led to a global environmental problem that calls for prompt interventions. Traffic congestion also affects the economic situation due to increased fuel consumption as a result of the lost time that the driver spends while stuck up, and due to delay in the arrival of goods on time due to the number of hours that are wasted due to traffic jam.

In the last decade, several researchers target finding scientific solutions to reduce traffic congestion. Most of the scientific studies take into account the dynamic attributes for road jams to build their own transportation management systems \cite{2}. In this context, Artificial intelligence has held the largest share through the implementation of decision Trees, fuzzy Logic and neural networks \cite{3}. But the greatest weakness in these techniques is the ignorance of the interaction that actually exists between the dynamic attributes (human, environment, etc) during traffic. Some literature studies have succeeded in reducing traffic congestion after using multi-agent based intelligent transportation systems (ITS). Soares et al., in \cite{4} applied a framework that combines the characteristics of the Multi-Agent System (MAS), JADE and SUMO for the development of multi-agent traffic solutions in contemporary transportation systems. The Simulation of the proposed model proved to be an effective approach to set up traffic solutions in socio-technical systems after being compared with real traffic flow network.

Rezzai et al. in \cite{5} proposed an intelligent system for traffic management for the aim of road traffic mitigation. This model consisted of 2 layers: environment’s perception and traffic management. However, it focused on the case of an isolated intersection with no realistic data. Their main objective was to estimate the density of traffic using vehicle density detection sensors, and through the control of traffic signals based on delay time records. The simulation revealed a 10\% decrease in traffic congestion.

Later on, Piris et al. \cite{6} proposed an intelligent traffic light using multi-agents to reduce the vehicle trip duration in Cologne (Germany). In this study, three agents were used: traffic light management agent, traffic jam detection agent, and the agentin charge of controling the traffic lights at an intersection. Their MAS was based on TraCI tool in SUMO software. The results analysis of the proposed model proved its capability to improve existing solutions such as conventional traffic light management systems (static or dynamic) in terms of reduction of vehicle trip duration (13\%).

In \cite{7}, Abu Bakaret al. applied a new model based on Discrete Event Simulation to avoid road traffic congestion in Jalan Mahkota, Kuantan and Pahang (Malaysia). They applied three different scenarios (S1, S2 and S3) to find the best fitting scenario. The parameters of car waiting time, number of vehicles in queue and the traffic light’s utilization were selected for performance measurement. Upon analyzing the experimental results, S3 was found to be the best performing scenario that can reduce the traffic congestion by $1.34$ minutes.

Later on, Sumia and Ranga \cite{8} proposed an intelligent traffic management system for prioritizing emergency vehicle in a smart city. The aim of this study was to give priority to ambulances to find the shortest possible path and field a counter measure to skip the obstruction caused by the traffic light system by allowing a hack during operation. The authors conducted a simulation and compared it with two systems: EPCS and Green Wave based on CUP Carbon Simulator. Results showed that the proposed model registered $12$\% improvement over EPCS system and $17$\% over Green Wave.

Zhang et al., \cite{9} developed new MAS to control traffic in China. The goals of this system were to control the green and red signals ratio at multiple adjacent intersections in traffic network.  This system was based on genetic algorithm to establish a distributed urban traffic control system that can be continuously optimized.  The results showed that the proposed system improved the traffic capacity of the traffic flow by $50$\% .

On the other side, Tewari et al., in \cite{10} used multi-agent reinforcement learning across multiple traffic intersections to mitigate traffic congestion in Bengaluru (India).The multi-agent reinforcement learning reduced the average density and delay by $33$\% compared to Fixed Signal Timing.

The goal of this paper is to apply a microscopic simulation carried out in Saeb Salam roundabout in Beirut, which has some crossroad intersections that witnesses daily traffic congestions especially at peak times. The aim of our study is to set up, validate and experiment a simulation of new multi-agent system using V2I communication model of current traffic by using agent-based traffic simulator SUMO and Jade(Java Agent DEvelopment framework) through TraSMAPI (Traffic Simulation Manager Application Programming Interface).

Our study is built on foundations that give knowledge about how to implement traffic simulation models and analyze the obtained results to investigate road traffic in real time. Two scenarios have been presented, the first addresses the stream of emergency vehicles relying on the interaction between the emergency vehicle agent and the Traffic Control Room (Traffic Room Manager) by activating the dynamic traffic light Agent and evaluates the performance of the proposed scenario through the observations of traffic flow, travel time and waiting time. The second scenario considers traffic congestions at crossroad and roundabout intersections, and allows for communication between three agents: Intersection Management Agent (IMA), Vehicle Mobile Application Agent and Traffic Room Manager through ACL (Agent Communication Language) messages. The interpretation of the controlled capacity of vehicles in a pool at the intersection and the figured out waiting time for a normal vehicle explains the figured out mitigation and claims resolving the gridlock problem.

\section{Place of Study}
Traffic congestion in Lebanon has turned into a transportation crisis that is getting worse day after another. Beirut comprises almost all governmental centers, ministries and public administration as well as private companies, large facilities of different fields, educational centers, clubs, etc. This factor is more than enough to understand the fact that residents spend hours stranded in traffic jams that happens every day and not only during peak hours. But this is not the sole factor. The cost of housing is unbelievably high in the capital what prompts people to dwell away from the city center and in most cases outside the city in other provinces. All of them commute to the capital on a daily basis allowing for a large flow of traffic especially at the entrances. 

Moreover, the road network is deteriorated due to poor infrastructure or lack of maintenance or non-compliance to standard specifications when constructed or all of them. Such condition besides distracted and erratic driving; and violation of traffic laws cause the occurrence of traffic accidents what makes the scene even more complicated. Furthermore, the growing population and increasing reliance of private cars due to the absence of a reliable public transportation system and other main reasons that contributes to the problem of traffic congestion leading to bottlenecks and gridlocks. 

\begin{figure}[b] 
\centering
\includegraphics[width=0.475\textwidth]{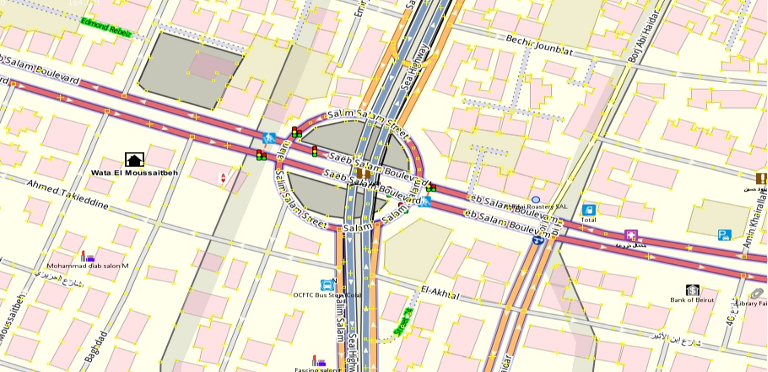}
\caption{Saeb Salam intersection Map} 
\label{fig1} 
\end{figure}

Our new model is built on Saeb Salam intersections in Beirut - Lebanon. The choice made on this junction for two reasons :First, in terms of traffic flow to and from it as this road falls on the south entrance to the city centre and because of having a lot of universities, public administrations, and facilities; and secondly, in terms of the presence of traffic lights interacting with Open Street Map (figure \ref{fig1}). This intersection is a huge roundabout that consists of three sub crossroads with active traffic lights; each consists of two or three lanes. 

\begin{figure}[h] 
 \centering
\includegraphics[width=0.475\textwidth]{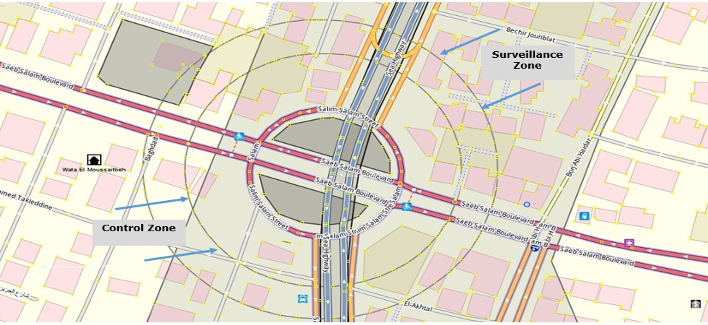}
\caption{Representation of Control and Surveillance zones for Saeb Salam Intersection} 
\label{fig2} 
\end{figure}

For the goal of defining the scope of communication and interaction between the various agents, we created two fictive concentric zones: \textbf{control zone} which is subject to IMA’s commands is about $0.3$ $Km^{2}$ (Radius $0.3$ $km$) and \textbf{surveillance zone} which is subject to IMA’s ceaseless monitoring is supposed about $0.8$ $Km^{2}$ (Radius $0.5$ $Km$) (see figure \ref{fig2}). \\
 
To make our experimental simulation more realistic, we are provided by the Traffic Management Center in Beirut with descriptive data relevant to Saeb Salam intersection. Such data report about the maximum number of vehicles that cross intersections at peak time (from 7:00 till 9:00AM and from 2:00 till 4:00 PM), and the average time that emergency vehicles take to cross the intersection controlled by static traffic lights. In addition, we were informed by the traffic control room that the static traffic lights in these intersections pass through six scheduled phases with different cycle length each as per the daytime (table \ref{tab1}). 
 
 \begin{table*}[]
 \begin{center} \footnotesize
\begin{tabular}{lp{0.15\textwidth}p{0.15\textwidth}p{0.15\textwidth}p{0.22\textwidth}}
 \textbf{Route ID} & \textbf{Junction Source}  & \textbf{Junction Destination} & \textbf{Lane Length (m)}  & \textbf{Maximum Vehicle Occupancy} \\ \hline
 1 & 393948024 & 6555914051 & 101.52 & 20 \\
 2 & 393948024 & 2356356625 & 223.94 & 44 \\
 3 & 287640809 & 287623849 & 136.96 & 27 \\
 4 & 393912045 & 2356356625 & 168.32 & 33\\ \hline
\end{tabular}
\end{center}
\caption{Maximum Vehicle Occupancy at the intersections}
\label{tab1}
\end{table*}

\section{Methodology}
In this study, the intelligent traffic management system assisted with the proposed communication protocol is developed to help vehicles drive across roundabout and crossroad intersections safely and efficiently.\\
In our model, the Traffic Management system consists of four Agents (see figure \ref{fig3}): 
\begin{figure}[h] 
\centering
\includegraphics[width=0.475\textwidth]{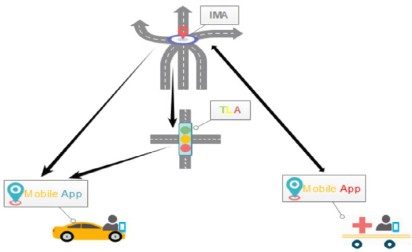}
\caption{Agents Interaction} 
\label{fig3} 
\end{figure}  
 \begin{enumerate}
 \item Intersection Management Agent (IMA) represents a targeted intersection Area 
 \item	Traffic Light Agent (TLAgent) represents the set of traffic lights in the targeted intersection area controlled by IMA 
 \item	Vehicle Agents represented by a mobile application (Mobile Agent) installed in the driver’s mobile controlled by IMA and TLA 
 \item	Traffic Room Manager Agent which is the management centre of all IMAs and it is located in Ministry 0f Interior and Municipalities 

 \end{enumerate}

Figure \ref{fig3} stands for the nature of agent’s interactions, as the IMA is responsible of a certain area that represents a crossroad or roundabout intersection, and allowed to get the occupation information at each intersection of his scope on real time basis. Each IMA is connected to TLA and Mobile Agent, and in mean time IMA is in direct contact with traffic room manager Agent.\\
The function of the traffic light agents, if dynamic mode is enabled, is to provide a well-coordinated sequence of traffic control plans using the traffic lights. Thus, the traffic light agents can be programmed to learn the behaviour that maximizes reward.

\subsection{Tools and Implementation}
Traffic lights at intersections and pedestrian crossings control the flow of traffic. Seeking the best performing transportation system, civilized countries developed and implemented intelligent dynamic road signs. Such systems can be adapted to the information received from a central unit about the position, speed and direction of vehicles, status of roads. They try to communicate with vehicles to alert drivers of impending light changes and thus reduce motorists' waiting time considerably. Trials are currently being conducted for the implementation of these advanced traffic lights but there are still many hurdles to widespread use that need to be addressed; one of which is the fact that few cars yet have the required systems to communicate with these lights.\\
Our proposed model consists of two scenarios, the first intends to reduce the delay time for emergency vehicles during a peak time and compare simulation results with the data collected at the intersection of Saeb Salam. As for the second, it aims to effectively ease road congestion at intersection, that is to say, avoiding gridlock intersection.\\

In order to implement these two scenarios, there shall be tools for creating agents and conduct simulations. So the option fell to use SUMO (Simulation of Urban Mobility) as a traffic simulator, JADE (Java Agent DEvelopment Framework) to develop agent-based applications, through TraSMAPI (Traffic Simulation Manager Application Programming Interface).

\subsubsection{Emergency Vehicle Scenario(Scenario-A)}
\begin{figure}[h] 
\centering
\includegraphics[width=0.475\textwidth]{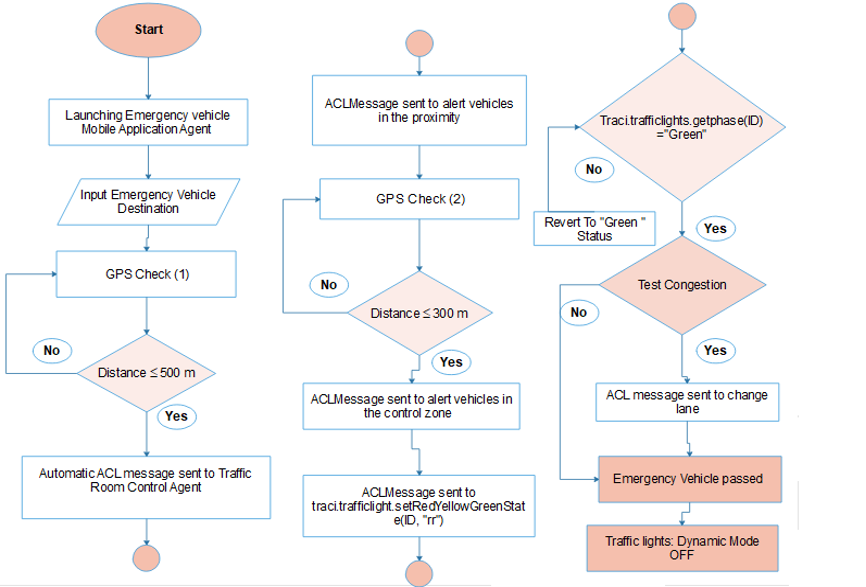}
\caption{Emergency vehicle Scenario (Scenario-A)} 
\label{fig4} 
\end{figure}

Initially, before going through a detailed explanation of the emergency vehicle scenario, it is necessary to recall the specifications of this type of vehicles. It is characterized by its speed for the privacy of its work compared to a normal vehicle.\\
Emergency vehicles, in their general sense, are vehicles that are concerned with responding to all life-threatening emergencies, for example, ambulances and firefighters; or responding to car failure emergencies. These vehicles are usually operated by designated agencies, often part of the government, but also run by charities, non-governmental organizations and some commercial companies.\\
For this scenario, depicted by figure \ref{fig4}, Emergency vehicle (Ambulance, Fire fighting truck…) cannot stop. In fact, every emergency vehicle has a mobile application interface different than that of any regular vehicle. Emergency vehicle knows exactly its desired destination and report it to Traffic Room Manager Agent while activating its application. 

This flow chart represents the means of continuously crossing emergency vehicle across different routes linked to the Saeb Salam roundabout in Beirut, which is characterized by several intersections with static traffic lights parked at its entrances.\\
Initially, once the driver operates the emergency vehicle, he launches its mobile application agent “Mobile Agent” to determine its position via GPS and by ACL message informs the traffic control room agent (Traffic Room Manager) with a continuous verification to detect when entering the Surveillance and Control zones. When this vehicle enters the surveillance zone, the Traffic Room Control agent sends an alert message by using ACL message informing all vehicles located in the control zone about the presence of an emergency vehicle on the outskirts of their zone.\\
Inside the control zone, the traffic lights phases at the intersection specified by the IMA become important, which obliges the traffic light agent “Tl Agent” to follow the instructions of the traffic control room agent, whether shifting to red or green. Thus, the traffic control room agent sends a notice to all vehicles in the same area that the traffic lights for roads that are not cleared by the emergency vehicle will turn red, so that this vehicle moves safely toward its destination without interruption. If this vehicle may encounter a traffic jam, the traffic control room agent informs to the driver via mobile application to change the lane to continue its trajectory without causing any additional delays. After the emergency vehicle being passed over the junction, traffic light dynamic mode is turned off which allows reverting back to the scheduled sequential phases. It should be noted that this scenario was applied to four different routes (route ID= “1, 2, 3, 4”) in the inspected area of Saeb Salam.\\
Furthermore, one of the important components of the health care system is the pre-hospital emergency care system. This section of the health system is characterized by heavy traffic, an increase in the urban population and an increased demand for emergency services before hospitalization. One way to solve this problem is to use ambulances.\\
For this reason, in the interface of the ordinary driver's mobile application, there is a background to report an emergency situation in the intersection area and the driver can upload an image that navigates in the case (accidents, etc ...). IMA communicates with the traffic control room to analyse the image and inform the nearest hospital or others to quickly call for urgent intervention and dispatch to the spot.

\subsubsection{Congestion and Gridlock Scenario (Scenario-B)}
\begin{figure}[htp] 
\centering
\includegraphics[width=0.475\textwidth]{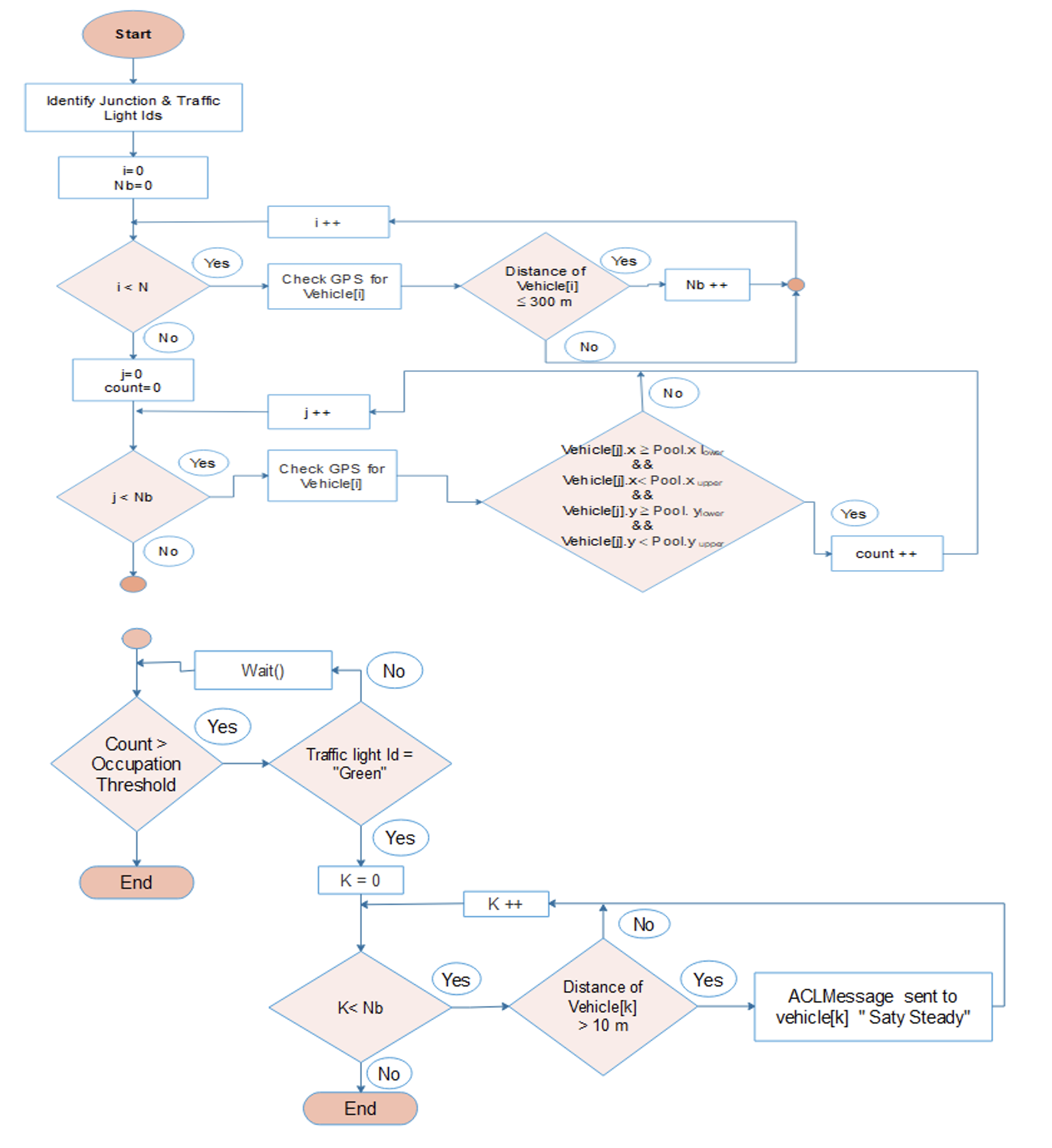}
\caption{Grid Lock Scenario (scenario-B)} 
\label{fig5} 
\end{figure}

Road congestion is known as one of the most serious crises facing nations that request prompt actions. The idea of this scenario aims to solve the congestion crisis and avoid gridlocks at intersections, because it is likely that it is the result of differences in speed between various on-road vehicles or traffic accidents or sudden breakdowns of vehicles in the lanes connecting the intersections with each other. This scenario passes through several stages, the driver of which must first activate his application to inform the traffic control room via a specialized Mobile Application Agent in order to include in the private database tracked by the Intersection Management Agent.\\
The traffic control room is assigned of identifying the IDs for each of the traffic lights and junctions under his scope, and generating statistics on the number of vehicles that cross the range of the control area when they reach an equal distance or less than 300 meters subject to continuous checking using the GPS system. The identification of vehicles’ ID is performed by checking the database that is the Register Log that records all the vehicles registered in distinct zones under the monitoring of Traffic Room Manager and comparing with the activated vehicle mobile application.
The mobile application agent automatically reports to the IMA via a form of ACL message informing that the vehicle has become within the range specified for the control zone. This application continues with a distance checking in order to be fully aware whether the vehicle has crossed the intersection zone by comparing its position coordinates with the lower and upper bounds of the intersection node, that we named pool.\\
In order to maximize traffic throughput and to better utilize the capacity of the intersection pool, the IMA needs to manage the space-time occupancies based on vehicle dynamics and live shared location. The IMA keeps tracking of the number of vehicles in the pool and compares it with an occupation threshold which is assumed equivalent to half the maximum number of vehicles can be occupied by the intersection pool. The assumption was made after several simulations to avoid falling into pool saturation and thus falling into a traffic congestion that gets more complicated when the traffic light of the other routes turns green and allows more vehicles going into the congested pool.\\
As the latitude and longitude of each vehicle in the control zone is already known, the number of cars coming in into pool is counted on current time basis, and thus obtaining real pool occupation. As the occupation exceeds occupation threshold, a pop message will be sent by IMA to all the vehicles that are in the control zone, but not those which are at less than 10m distance from the pool, to keep steady even the light is GREEN. That is when the light goes RED on the said route and green on other lane as scheduled, the car flow there will go smoothly without any interruptions. The assumption that the ACL message is sent to those exceeding 10 meters from the pool exclusively for the fact that the vehicles that lie within the first two rows from the pool’s edge and responding to the current green traffic light can do nothing with the coming ACL message and even this may cause traffic accidents (see figure \ref{fig5}). 

\section{Simulation Results }
\begin{figure}[htp] 
\centering
\includegraphics[width=0.475\textwidth]{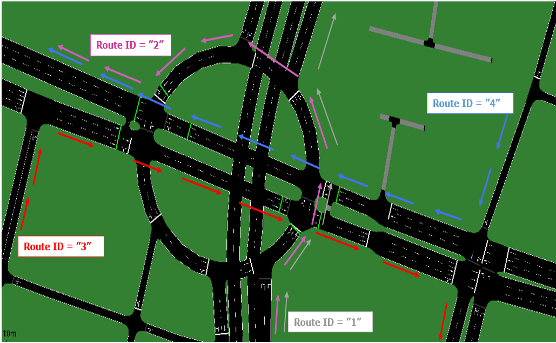}
\caption{Representation of routes} 
\label{fig6} 
\end{figure}

\begin{figure}[htp] 
\centering
\includegraphics[width=0.475\textwidth]{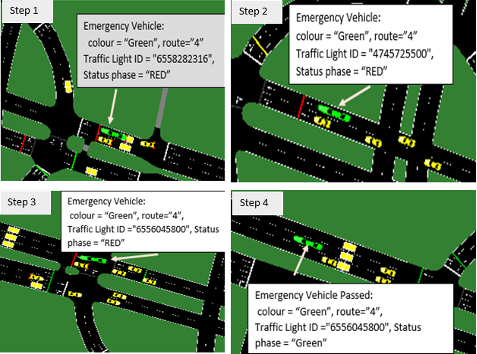}
\caption{Emergency Vehicle crossing the route “4” –Current Case} 
\label{fig7} 
\end{figure}

The emergency vehicle simulation model was implemented on the four different specific routes (routes: “1, 2, 3, 4”) connected to the intersection at Saeb Salam, based on the representation below (figure \ref{fig6}).  Each route has an identifier route ID and a list of edges that cross it.
To represent a realistic model through Sumo, we have defined two types of vehicles “Passenger” and “Emergency” which have several characteristics, such as length, type, the maximum speed, etc.\\
Figure \ref{fig7} shows the simulation of an emergency vehicle stream for a current case described by 4 steps that demonstrate its passage across 3 junctions, each having a traffic light with distinct ID. We performed several simulations for the current case to calculate the averages of speed, acceleration, time Loss and travel time. It should be noted here that the time Loss denotes the average time wasted due to driving slower than desired and includes Waiting Time (in seconds). It can be obtained by using the function Sumo Vehicle. Get Time Loss; while the travel time is the time taken by the vehicle to arrive to their destination and is computed using the formula: Time Travel= Speed/ average acceleration.
\begin{equation}
Time\;Travel= Speed/ average\;acceleration
\end{equation}

\begin{figure}[htp] 
\centering
\includegraphics[width=0.475\textwidth]{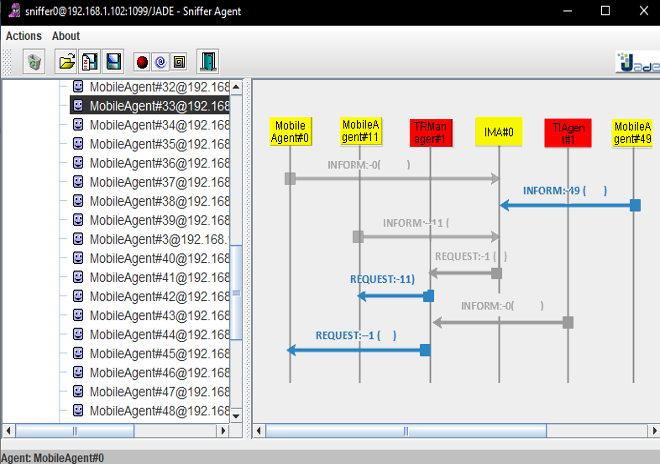}
\caption{Representation of Agents Communication} 
\label{fig8} 
\end{figure}

The application of this model aims to compare measurements such as speed, acceleration, time Loss and travel time averages of emergency vehicles between the current case at Saeb Salam and the proposed model. The construction of the new model within the intersections of Saeb Salam had to take into account the same criteria and standards in order to prove its qualification.\\
The following illustration (figure \ref{fig8}) shows that the emergency vehicle crossed all the intersections following the planned route after declaring it to the traffic control room manager, and its passage completed without any interruption. \\

We placed detectors on each road in two streets or three, at a distance of 300 meters from the first traffic light entering the pool and at 300m after the last signal exiting the pool relevant to the route targeted by the emergency vehicle.

When the vehicle passes the detector specified at the entrance, this calls the IMA to send an alert message to all vehicles in the proximity that an emergency vehicle is about to pass and that the signals on routes 1 and A for example will turn red. A communication intra-agent was represented by Sniffer agent using ACL messages; the representation in figure \ref{fig7} presents a sample of communication held among agents.

Firstly, MobileAgent\#0, MobileAgent\#49 and Mobile Agent\#11 reported to IMA by triggering the mobile application. Next, IMA sent a request message to TR Manager\#1 reporting that Mobile Agent\#49 is an emergency vehicle and just entered the control zone whereas TRManager\#1 in turn delivered an alert message to other vehicles in the control zone informing them that there is an emergency vehicle tending to take this way: please “stay Steady”. Also TLAgent\#1 reported to TR Manager\#1 about its phase status. 

\begin{figure}[htp] 
\centering
\includegraphics[width=0.475\textwidth]{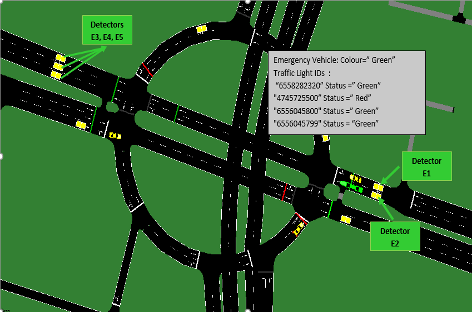}
\caption{Emergency Vehicle Simulation in Scenario-A case} 
\label{fig9} 
\end{figure}

The simulation of emergency vehicle in the Scenario-A case (applying the new proposed scenario) was done by using the tool-chain that consists of JADE and SUMO-GUI. The result of several simulations showed non-stop emergency vehicle in the route “4”, where the traffic lights in each junction of its path has a green status which allows its smooth passage across the junctions without interruptions (see figure \ref{fig9}).

\subsection{Gridlock Simulation Model}
\begin{figure}[htp] 
\centering
\includegraphics[width=0.475\textwidth]{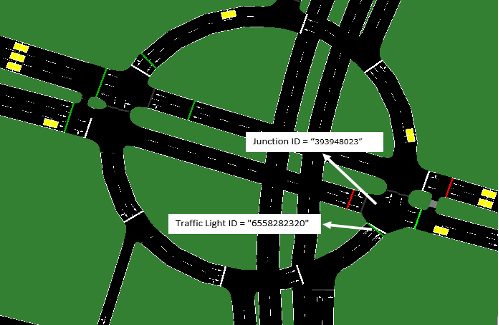}
\caption{EJunction ID= “393948023” representation} 
\label{fig10} 
\end{figure}

As mentioned before, the gridlock simulation model contributes to solving the problem of gridlocks at intersections. It was built on the basis of communication among the four agents; and its evaluation was done upon comparison between the current model and Scenario-B model (applying the new proposed scenario for gridlocks). This study was conducted on a single junction the Saeb Salam roundabout (Junction ID = “393948023”). The junction under study is illustrated in the figure below (figure \ref{fig10}).

\begin{figure}[htp] 
\centering
\includegraphics[width=0.475\textwidth]{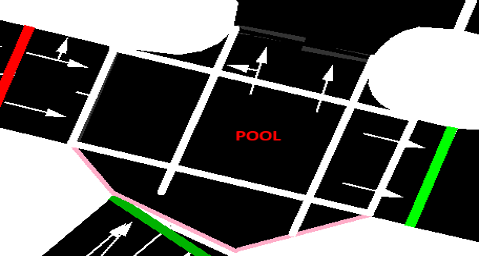}
\caption{Representation of the Pool} 
\label{fig11} 
\end{figure}

We assume that the inter-shape of the junction’s pool is of square shape ($Area= 6.4\times 6.4= 40.96$ $m^{2}$), so the pool has a maximum vehicle occupancy of 5 vehicles. Figure \ref{fig11} illustrates the junction’s pool. 

For the current case, the simulation model was based on an estimate of a number of vehicles likely to cross the aforementioned intersection, and it is common to have the number of vehicles exceeding the maximum load capacity of this area what end up in critical gridlocks. This simulation aims to deal with issue that arouse the driver's interest in terms of getting late to work, for example, or goods delivery delay. Most of these issues are related to the time factor which has an effective impact on our daily life and its success. We had to get relevant criteria by applying several methods to java classes.

\begin{figure}[htp] 
\centering
\includegraphics[width=0.475\textwidth]{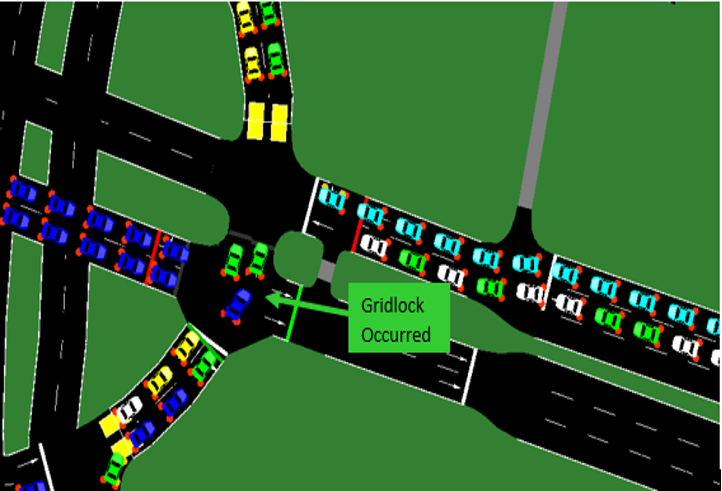}
\caption{Gridlock occurred at the junction ID=” 393948023”} 
\label{fig12} 
\end{figure}

We consider the following criteria: the averages of speed, acceleration, time Loss and travel time. The figure below shows that a gridlock occurred when there are 3 vehicles waiting in the pool (greater than occupation threshold in the determined junction), unable to cross the pool to the opposite side due to the traffic jam there, and thus exceeding its carrying capacity; and the traffic light status is still green. The vehicles which are allowed to pass by green light will keep on getting into the pool until its complete saturation. When the traffic light turns red and the vehicles in the left route are now allowed to move, they will get stuck in the saturated pool causing a large time loss for the vehicles in all the routes linked to the said junction (figure \ref{fig12}).

This situation affects the periphery of this junction on all sides, and could last for hours. Such case happens every day in Beirut and particularly in the peak hours. Here appears the need for a new approach with new concepts and features of Intelligent Transportation Systems, but in the same time takes into account the economic challenges of Lebanon, the developing country. 

This new model was built on stages similar to the previous model in terms of creating the same agents and map (Saeb Salam) but we included the identification of the studied junction(ID=”393948023”).

 For the simulation of the Scenario-B model, we put 6 detectors to count the number of vehicles (at the pool’s Entry and Exit), and we tested the counting number inside the pool by using a series of “if-condition”. When the total number reaches half the maximum occupancy which is in our case study equivalent to 2.5 assumed 2, traffic room control sends a request message for vehicles located at 10 meters and beyond away from the traffic light, with a notification to “Stay Steady “although the light is green. So even if the said traffic light turns red and that of the left route turns green, those cars of the left route receiving the same message are not allowed to move forward until the count is below the occupation threshold and another message is received to move if it’s green. Here it is necessary to draw attention that upon running the simulation while a vehicle is in Steady Mode. This time is added to the time loss.
 
\begin{figure}[h] 
\centering
\includegraphics[width=0.475\textwidth]{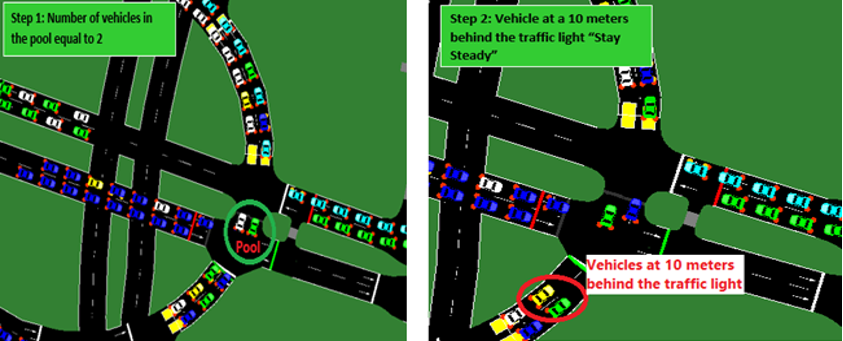}
\caption{Steps of gridlock simulation model} 
\label{fig13} 
\end{figure}

The representation below (figure \ref{fig13}) shows 2 steps of the simulation, the first when the number of cars in the pool is equal to occupation threshold, and the second step the vehicles that lie in the 3rd row and the rows that follow in steady mode... This operation was done through a function of Vehicle. Set Stopped (VehID) by identifying the vehicle ID and corresponding position. Such parameters were obtained by using other functions such as Sumo $Vehicle.getID ()$.

\section{Results and Discussion}
The results of the two new simulation models in our study referring to the two proposed scenarios are shown in Table \ref{tab2} \& Table \ref{tab4}. A comparison based on certain criteria related to the time factor is held for each scenario between the current case and the scenario case. As for the first simulation model, it addressed the stream of an emergency vehicle without stopping from its source to its destination, passing through two or three intersections, depending on the route it took. The results were based on operations repeated more than 50 times, so in each operation we had to measure the criteria related to the emergency vehicle such as the speed rate, the acceleration rate, the access time and the Travel time. This method was activated on the four different available routes, in both current and Scenario-A cases. The results of the simulation on the four routes were concluded for the two cases in Table \ref{tab2}.

\begin{table*}[]
\footnotesize
\begin{tabular}{|l|l|l|p{1.3cm}|p{1.3cm}|p{1.1cm}|p{1.8cm}|p{1.1cm}|p{1.1cm}|p{1.3cm}|p{1.1cm}|p{1.8cm}|p{1.1cm}|p{1.1cm}|}
\hline
\multicolumn{3}{|l|}{\parbox{2cm}{Routes}} &        & \multicolumn{10}{l|}{Average of simulation parameters}                     \\ \hline
\multirow{2}{*}{ID} &
  \multirow{2}{*}{\parbox{1.7cm}{Junction Source}} &
  \multirow{2}{*}{\parbox{1.7cm}{Junction Destination}} &
   &
  \multicolumn{5}{l|}{Current case} &
  \multicolumn{5}{l|}{Scenario-A case} \\ \cline{4-14} 
 &
   &
   &
  Lane length ($m$) &
  Number of vehicle &
  Speed ($m/s$) &
  Acceleration ($m/s^{2}$) &
  Time loss ($s$) &
  Travel time ($s$) &
  Number of vehicle &
  Speed ($m/s$) &
  Acceleration ($m/s^{2}$) &
  Time loss ($s$) &
  Travel time ($s$) \\ \hline
1  & 393948024  & 6555914051 & 101.52 & 20 & 11.38 & 0.6   & 41.16  & 59.96  & 20 & 11.14 & 0.57  & 19.99  & 39.53 \\ \hline
2  & 393948024  & 2356356625 & 223.94 & 44 & 19.33 & 0.586 & 78.991 & 111.97 & 44 & 20.05 & 0.579 & 38.997 & 73.62 \\ \hline
3  & 287640809  & 2876223849 & 136.96 & 27 & 9.2   & 0.53  & 72.04  & 90.29  & 27 & 10.23 & 0.59  & 30.43  & 47.76 \\ \hline
4  & 393912045  & 2356356625 & 168.32 & 33 & 10.02 & 0.61  & 58.96  & 75.386 & 33 & 12.3  & 0.59  & 29.13  & 49.97 \\ \hline
\end{tabular}
\caption{Simulation Results of Average of Parameters (Emergency Vehicle Model)}
\label{tab2}
\end{table*}

\begin{table*}[]
\begin{tabular}{|l|p{2.4cm}|p{1.4cm}|p{1.2cm}|p{1.9cm}|p{1.2cm}|p{1.2cm}|p{1.3cm}|p{1.2cm}|p{1.9cm}|p{1.2cm}|p{1.2cm}|}
\hline
Junction                     &   & \multicolumn{10}{l|}{Average simulation parameters}                      \\ \hline
\multirow{2}{*}{Junction ID} &   & \multicolumn{5}{l|}{Current case} & \multicolumn{5}{l|}{Scenario-B case} \\ \cline{2-12} 
 &
  Maximum vehicle occupancy &
  Number of vehicles &
  Speed ($m/s$) &
  Acceleration ($m/s^{2}$) &
  Time loss ($s$) &
  Travel time ($s$) &
  Number of vehicles &
  Speed ($m/s$) &
  Acceleration ($m/s^{2}$) &
  Time loss ($s$) &
  Travel time ($s$) \\ \hline
393948023                    & 5 & 3 & 12.3 & 0.74 & 451.32 & 467.94 & 2  & 14.25  & 0.42 & 222.79 & 256.59 \\ \hline
\end{tabular}
\caption{Simulation Results of Average of Parameters (Gridlock Model)}
\label{tab4}
\end{table*}


The analysis of these results requires recalling a brief definition of each of the used criteria of our study, where the Time Loss (in second) represents theaverage time lost due to driving slower than desired including Waiting Time; and the travel time (in second) is the time of the entire journey from the source to the planned destination including the Time Loss. \\

Firstly, these results showed that on the route ID- “1”, there was a 51.43\% reduction in the average Time Loss in the Scenario-A case compared to the average Time Loss in the current case based on the following formula:

\begin{multline}
\scriptsize
Reduction\; in\; Time \;Loss (\%)=\\ \frac{(Time\; Loss\; in\; Current\; Case - \;Time\; Loss\; in\; Scenario\; A\; case )}{(Time\; Loss\; in \;Current\; Case) } \\ \times 100
\end{multline}

We got similar percentages for the second and fourth routes (route ID=”2” and “4”). As for the third route (route Id =”3), the Time Loss was reduced by 57.75\%.Reductions in Time Loss and Travel Time (\%) upon applying Scenario-A for the four routes are listed in Table \ref{tab3}.

\begin{table}[h]
\footnotesize
\begin{tabular}{p{2.1cm}p{3.4cm}p{3.4cm}}
\textbf{Route ID}		& \textbf{Reduction in Time Loss \%}  & \textbf{Reduction in Travel Time \%}  \\ \hline
 1 & 51.43 & 34.07 \\
2 & 50.63 &  33.98\\
3  & 57.76 &  47.1\\
 4 & 50.59 & 33.71 \\ \hline
\end{tabular}
\caption{Percentage Reduction in Time Loss \& Travel Time upon applying Scenario-A
}
\label{tab3}
\end{table}
Talking about Travel Time, the reduction is ranged between 33\% and 47\%.\\

Based on these results, we can state that the hypothetical New World model based on multiple agents reduced the average Loss time by 52.6\% and the average Journey Time by 37.21\%. \\

The same context that was applied in the first model in terms of allowing multiple operation of the simulation model in both cases was also applied to the model of gridlocks (Scenario-B). Table \ref{tab4} reports the recorded results.\\

Reductions in Time Loss and Travel Time (\%) upon applying Scenario-B are listed in Table \ref{tab5}:
\begin{table}[h]
\footnotesize
\begin{tabular}{p{2.1cm}p{3.4cm}p{3.4cm}}
\textbf{Junction ID}		& \textbf{Reduction in Time Loss \%}  & \textbf{Reduction in Travel Time \%}  \\ \hline
 393948023 & 50.64 & 45.17 \\ \hline
\end{tabular}
\caption{Percentage Reduction in Time Loss \& Travel Time upon applying Scenario-B}
\label{tab5}
\end{table}

Based on the above, the significance of Gridlocks Model was clearly inferred as the average journey time of the vehicles crossing the said Junction was reduced by 45\% and the occurrence of gridlock was avoided.  

\section{Conclusion}
SUMO and JADE tool-chain is a new world of evolution in the simulation of intelligent traffic systems. The worsening situation of traffic congestion and the gridlocks at intersections have their negative impacts on human health, productivity, environment, economics, etc. Our research proposed new 2-scenarios, agent-based model that benefits from the interaction of its four agents to address the crisis of traffic congestion at the Intersection of Saeb Salam (Beirut), which contains several crossroad intersections. Each agent was assigned certain tasks and decisions were made based on the information exchanged among the four agents (Traffic Room Manager Agent, Intersection Manager Agent, Traffic Light Agent and Vehicle Mobile App Agent). There were two scenarios that constituted the new model. The first scenario (Scenario-A) studied the stream of Emergency Vehicles and proposed a solution that ensured the fastest and safest journey without interruptions. The second scenario (Scenario-B) addressed and figured-out effective and efficient solution for the issue of gridlocks at intersection. \\

The simulation results of both scenarios were compared with the simulation results of the current case that we were properly notified by the Traffic Room about all related current observations, statistics and traffic lights scheduled cycles. Upon interpreting the obtained results, we found out that for Scenario-A, the average Time Loss was reduced by 52\% and the average of the entire Journey Time by 37\% which means that emergency services can be completed faster and safer. As for Scenario-B, the average Time Loss was reduced by 50\% and the average of the entire Journey Time by 45\% which is a real enhancement that reflects smoother daily life, higher throughput, happier citizens and better community. It is worth mentioning that the proposed model suits developing countries with many challenges such as Lebanon as it is characterized by efficiency, low cost, flexibility and reliability.

\begin{acknowledgments}
This work was supported by the NSF, DOE, AFOSR, and ARO. SB and SY acknowledge support from the NSF GRFP. LA acknowledges support from the HQI. EC acknowledges support from the NRF of Korea (2021R1C1C1009450, 2020R1A4A1018015)
\end{acknowledgments}


\begin{thebibliography}{10}
\bibitem{1}
{\sc V. Lomendra, P. Sharmila, D. Ganess, N. Vandisha}, {\em Assessing the causes \& impacts of traffic congestion on the society, economy and individual: a case of Mauritius as an emerging economy}, Studies in Business and Economics, Vol.13, issue 3, 2018, pp: 230-240.

\bibitem{2}
{\sc W. Wang , R. Guo and J. Yu}, {\em Research on road traffic congestion index based on comprehensive parameters: Taking Dalian city as an example}, Advances in Mechanical Engineering, Vol. 10, Issue 6, 2019,  pp: 1-8.

\bibitem{3}
{\sc S.Al-Garni and A. benAbdennour}, {\em A Neural Network Based Traffic Flow Evaluation System for Highways}, Journal of King Saud University - Engineering Sciences, Vol. 20, Issue 1, 2008, pp: 37-45.

\bibitem{4}
{\sc G. Soares, Z.Kokkinogenis, J.Macedo, R.Rossetti}, {\em Agent-Based Traffic Simulation Using SUMO and JADE: An Integrated Platform for Artificial Transportation Systems}, Simulation of Urban MObility User Conference, 2014.

\bibitem{5}
{\sc M. Rezzai, W.  Dachry, F. Moutaouakkil, H. Medromi}, {\em Designing an Intelligent System for Traffic Management}, Journal of Communication and Computer, 12(3), 2015, pp. 106-114.

\bibitem{6}
{\sc L. Piris, D. Rivera, I. Maestre, E. de la Hoz , S. Fernandez}, {\em Intelligent Traffic Light Management using Multi-Behavioral Agents}, Actas de las XIII Jornadas de Ingeniería Telemática, 2017, pp: 110-117.

\bibitem{7}
{\sc N. A. A. Bakar, A. F. N. Adi, M. A. Majid, K. Adam, Y. M. Younis and M. Fakhreldin}, {\em The Simulation on Vehicular Traffic Congestion Using Discrete Event Simulation (DES): A Case Study}, 2018 International Conference on Innovation and Intelligence for Informatics, Computing, and Technologies (3ICT), Sakhier, Bahrain, 2018, pp. 1-6.

\bibitem{8}
{\sc L. Sumi and V. Ranga}, {\em Intelligent traffic management system for prioritizing emergency vehicles in a smart city (technical note)}, International Journal of Engineering, Transactions B: Applications, vol.31, no.2, 2018, pp.278-283.

\bibitem{9}
{\sc Z. Zhang, J. Ye, S. Cheng}, {\em Research on Traffic Congestion Resolution Mechanism based on Genetic Algorithm and Multi-Agent}, 3rd International Conference on Mechatronics Engineering and Information Technology (ICMEIT 2019), 2019.

\bibitem{10}
{\sc U.Tewari , , V. Raveendran, V.Sudhakaran}, {\em  Intelligent Coordination among Multiple Traffic Intersections Using Multi-Agent Reinforcement Learning}, 33rd Conference on Neural Information Processing Systems, Canada, 2019, pp : 1-10. 

\end{thebibliography}
\end{document}